\documentclass[review]{elsarticle}
\newcommand{\ba}{\backslash}
\usepackage{lineno,hyperref,bm,amsfonts,graphicx,amsmath,color,booktabs}
\usepackage{algorithm}
\usepackage{algpseudocode,subfigure}
\modulolinenumbers[5]

\begin{document}

\begin{frontmatter}

\title{Target observation of complex networks}


\author[mymainaddress,mysecondaryaddress]{Yi-Fan Sun\corref{mycorrespondingauthor}}
\cortext[mycorrespondingauthor]{Corresponding author}
\ead{sunyifan@ruc.edu.cn}

\author[mysecondaryaddress]{Zheng-Yang Sun}

\address[mymainaddress]{Center of Applied Statistics, Renmin University of China, Zhong-Guan-Cun Street 59, Beijing 100872, China}
\address[mysecondaryaddress]{School of Statistics, Renmin University of China, 
Zhong-Guan-Cun Street 59, Beijing 100872, China}

\begin{abstract}
How to observe the state of a network from a limited number of measurements has become an important issue in complex networks, engineering, communication, epidemiology, etc. Under some scenarios, it is either unfeasible or unnecessary to observe the entire network. Therefore, we investigate the target observation of a network in this paper. We propose a target minimal dominating set problem corresponding to target observation, which is a natural generalization of classical minimal dominating set problem. Three algorithms are proposed to approximate the minimum set of occupied nodes sufficient for target observation. Extensive numerical results on computer-generated random networks and real-world networks demonstrate that the proposed algorithms offer superior performance in identification of a target minimal dominating set. 
\end{abstract}


\end{frontmatter}


\section{Introduction}
How to efficiently detect, and then control, the state of a networked dynamical systems has been arising as a central issue in the context of complex networks \cite{Liu2011,Liu2013,Stefan2014Controllability,Gates2015Control, Liu2016Control,Haber2018State,Tzoumas2018Selecting}. In some real-world systems, e.g., power-grid networks \cite{Yang}, wireless networks \cite{Islam2009Distributed}, and mobile \emph{ad hoc} networks \cite{Cokuslu2006}, their states can be monitored in real time by placing sensor devices on the nodes. Because a sensor device is generally able to observe not only the state of the residing node but also all of its adjacent nodes in the network, as in the case of power grids \cite{Phadke1993Synchronized}, we only need to place sensors on a limited number of nodes. These occupied nodes thus constitute a \emph{dominating set}  (DS) of the network. The nonzero cost incurred by placing a sensor prompts us to seek for a DS with the lowest cost. Suppose the cost of placing a sensor is the same for all nodes. Then the problem is converted into finding the \emph{minimum dominating set} (MDS), i.e., the smallest subset of nodes such that each node of the network either belongs to this set or is adjacent to at least one node in this set. The MDS problem is a classical nonderterministic polynomial-time hard problem \cite{Garey} in graph theory, and there is a vast literature on it \cite{Haynes}. The fields of mathematics and theoretical computer science focus on constructing various bounds on the size of the MDS and quantifying the complexity of solving MDS problem \cite{Henning2009,Hansberg2010}. The field of physics, instead, pays more attention to analyzing the statistical properties of the MDS on various networks and designing heuristic algorithms to search for a near-optimal solution \cite{Zhao2014Statistical, Sun2016,Nacher2012Dominating,F2013Minimum,Takaguchi2014Suppressing}. 

In practice, however, we often witness some large technological, biological, and social systems. Their huge size and high complexity make observing the entire system unfeasible and unnecessary. Under some scenarios, instead, it is sufficient to target observe the system, that is, to observe the states of a preselected subset of nodes, called \emph{target nodes}, that may play a vital role in maintaining the normal operation of the system \cite{Gao2014Target}. Some nodes need to be occupied to make all the target nodes observable, and the set of occupied nodes is called as \emph{target dominating set} (t-DS).   
The so-called \emph{target minimal dominating set}  (t-MDS) problem is to seek for a t-DS with the minimum size.  A closely related problem is the partial MDS (p-MDS) problem, which has been proposed already in the literature \cite{Case,Das}. The difference between them is that the t-MDS problem searches for a smallest subset of nodes to observe a specified subset of nodes, while the p-MDS problem focuses on detecting a set of nodes of minimum size to guarantee that at least some given fraction of nodes is observed. Note that if each node in the network becomes a target node, t-MDS problem reduces to the MDS problem. Thus t-MDS problem can be viewed as a generalization of MDS problem. 

Although target observation has many potential applications in various areas such as engineering \cite{Yang2008PI} and biology \cite{Stefan2014Controllability}, research from the algorithmic viewpoint on identification of t-MDS is scarce as far as the authors know. The approach most likely to be devised is to implement an algorithm to get a near-minimum DS first, and then remove the nodes from the DS whose removal will not make all target nodes unobservable. Another, perhaps easier, way is to extract the subnetwork composed of \emph{only} the target nodes and the edges between them and then seek a near-minimum DS of this subnetwork. Under many scenarios, however, these two approaches are not the optimum ones: that is, they overestimate the minimum number of nodes needed for target observation. 
A toy example is presented in Fig. 1. The network is composed of 9 nodes and 10 edges. Suppose we want to observe a subset of nodes, $\{2,7\}$. The first approach focuses on the construction of MDS for the full network, and thus it will occupy the nodes 2,3 and 7 that can observe as many nodes as possible. Because node $3$ is irrelevant to observation of the target nodes, it is deleted from the solution. Hence the first approach reports that we need to occupy at least two nodes for target observation [Fig. 1 (a)-(c)]. Because nodes 2 and 7 are not directly connected, the subnetwork induced by these two nodes consists of the two isolated nodes only. The second approach only considers occupying the target nodes, and thus it also predicts that we need to occupy two nodes, i.e., nodes 2 and 7, in the toy example. However, in reality, we only need to occupy one node, i.e., node 5 [Fig. 1 (d)]. This result prompts us to develop an algorithm to approximate the t-MDS sufficient for target observation. 

The rest of paper is organized as follows. In Section 2, we define the t-MDS problem in networks and introduce some notations. In Section 3, we present three solving algorithms--a message-passing-based algorithm and two local algorithms--to get a near-minimum t-DS for a single network. In Section 4, we apply the proposed algorithms to computer-generated random networks and several large-scale real-world networks and compare their performances with other alternatives. Finally, in Section 5, we summarize the main work of the paper. 

\begin{center}
\resizebox{0.9\textwidth}{!}
{\includegraphics{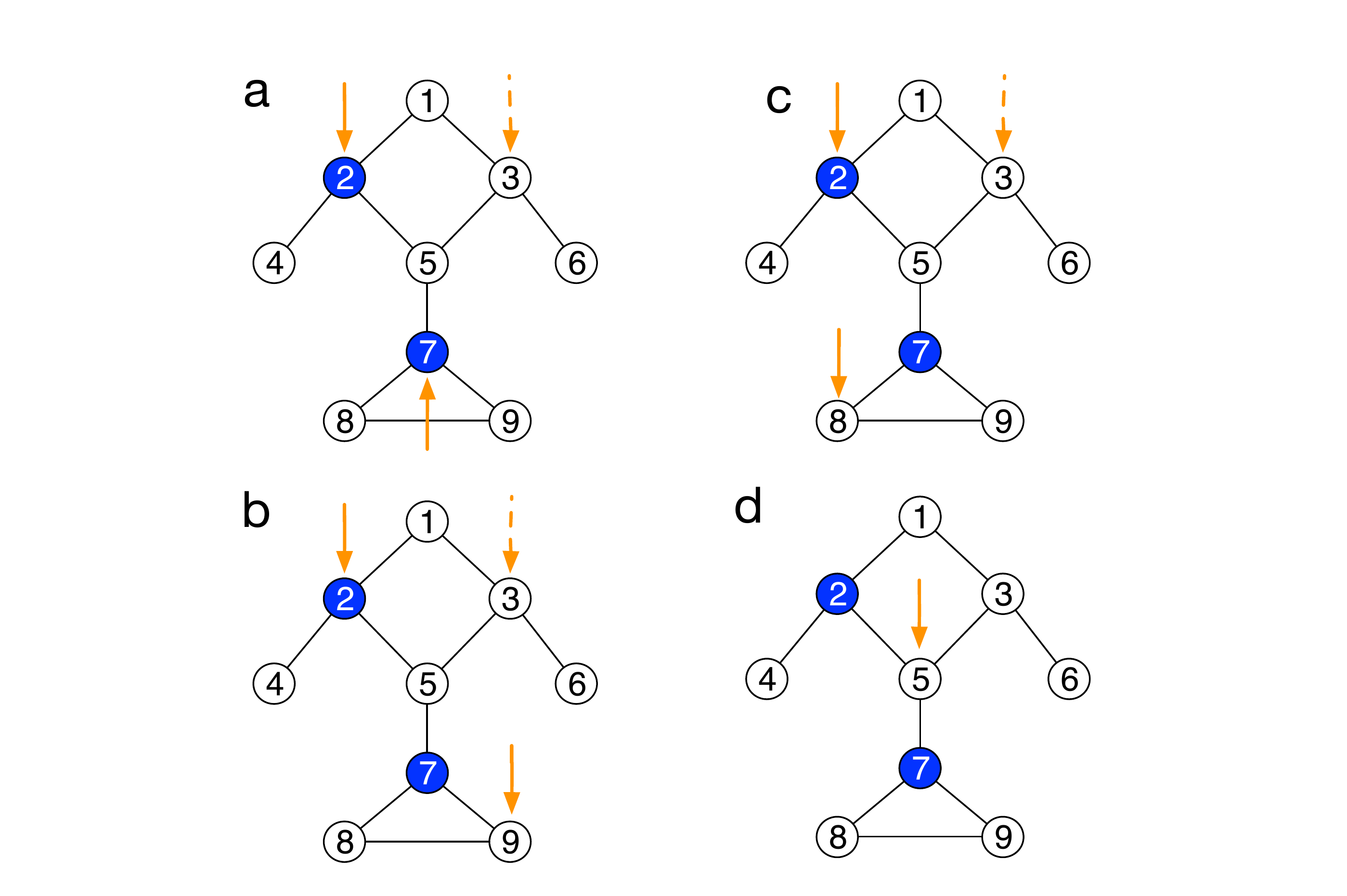}}\\[3pt]  
\label{fig:toy}
\parbox[c]{15.0cm}{\footnotesize{\bf Fig.~1.} (color online) t-MDS solutions for target observation of a toy network. We want to observe the target nodes 2 and 7. (a)-(c) show three t-MDS solutions by using the first approach. Each node in t-MDS is pointed by an arrow with solid line. The first approach first finds a MDS of the full network, which consists of three nodes (pointed by the arrows), and then deletes the node irrelevant to the target observation (pointed by the arrow with dashed line). Hence the first approach predicts that we need to occupy at least two nodes to realize the target observation. (d) shows the true t-MDS solution for target observation, which consists of only one node, i.e. node 5.}
\end{center}

\section{Definition of the t-MDS problem and notation}
Consider an undirected network $G=(V,E)$ composed of $N$ nodes and $M$ edges, where each edge connects between two distinct nodes. Two nodes $i, j\in V$ are \emph{adjacent}, if they are directed connected by an edge, and $j$ is called a \emph{first neighbor} (\emph{neighbor} for short) of $i$ and vice verse.  The \emph{degree} of a node $i$, denoted as $K_i$, is the total number of its neighbors. 
The \emph{mean degree} of network $G$, denoted as $\langle K \rangle$, is the average degree of nodes in $G$. 

Each node in the network $G$ is either occupied or unoccupied.  A node $i$ is observed if it is occupied ($v_i=1$) or adjacent to at least one occupied node. A set $\Gamma\subset V$ of nodes in network $G$ is a \emph{dominating set} (DS) if each node $i\in V$ is observed after occupying all nodes in $\Gamma$. The \emph{minimum dominating set} (MDS) problem, a classical NP-complete problem, is to determine a minimum size DS.

Under some scenarios, we only need to observe a subset $V_t\subset V$ of target nodes. For the set $V_t$ of target nodes, a target dominating set (t-DS) $\Gamma_t$  refers to a subset of nodes such that each node in $V_t$ is observed, that is, is either in the set $\Gamma_t$ or adjacent to at least one node in the set $\Gamma_t$. The \emph{target minimum dominating set} (t-MDS) problem entails searching for a minimum size t-DS, denoted as $\Gamma_t^{\star}$. Note that when $V_t=V$, i.e., all the nodes are the target nodes, t-MDS problem reduces to the MDS problem. Hence, the t-MDS problem can be seen as a generalization of MDS problem. 
  
If node $i$ is neither in set $V_t$ nor adjacent to at least one node in set $V_t$, node $i$ must not belong to $\Gamma_t^{\star}$ because occupying node $i$ is unable to make any node in $V_t$ observable. Let $\partial V_t$ represent the nodes not in $V_t$ but adjacent to at least one node in $V_t$. Therefore, $\Gamma_0$ must be a subset of $V_t\bigcup \partial V_t$. We define a subnetwork induced by set $V_t\bigcup \partial V_t$, denoted as $G_t$, as the nodes in $V_t$ and $\partial V_t$ along with the edges that links \emph{at least one node in $V_t$}. Be different from the simple subgraph of target nodes, the subnetwork $G_t$ defined here includes not only the target nodes but also their adjacent nodes. For example, in the toy example presented in Fig.~2(a), the simple subgraph of target nodes $V_t=\{2,7\}$ consists exclusively of the two nodes [Fig.~2 (c)], whereas the subnetwork $G_t$ also includes their adjacent nodes $\partial V_t=\{1,4,5,8,9\}$ [Fig.~2 (b)]. Note that each edge in $G_t$ connects with at least one node in $V_t$, and thus the edge between two nodes in $\partial V_t$, for instance, edge $(8,9)$ in the toy example, is  not included in $G_t$, suggesting that $G_t$ is a subset of the simple subgraph of $V_t\bigcup \partial V_t$ which include \emph{all} the edges between nodes in $V_t\bigcup \partial V_t$ [Fig. 2 (d)]. 

Obviously, the t-MDS $\Gamma_t^{\star}$ is also a t-DS of subnetwork $G_t$ for set $V_t$ and, furthermore, is the minimal one (because if subnetwork $G_t$ has a t-DS $\Gamma_t\subset \Gamma_t^{\star}$, $\Gamma_t$ must also be a t-DS for the full network $G$, which contradicts the fact that $\Gamma_t^{\star}$ is a t-MDS of network $G$). Similarly, each t-MDS of subnetwork $G_t$ is also a t-MDS of network $G$. Therefore, we conclude that the full network $G$ and the subnetwork $G_t$ have the same t-MDS. The t-MDS problem in network $G$ is, thus, transformed into the t-MDS problem in subnetwork $G_t$. 

\begin{center}
\resizebox{0.8\textwidth}{!}
{\includegraphics{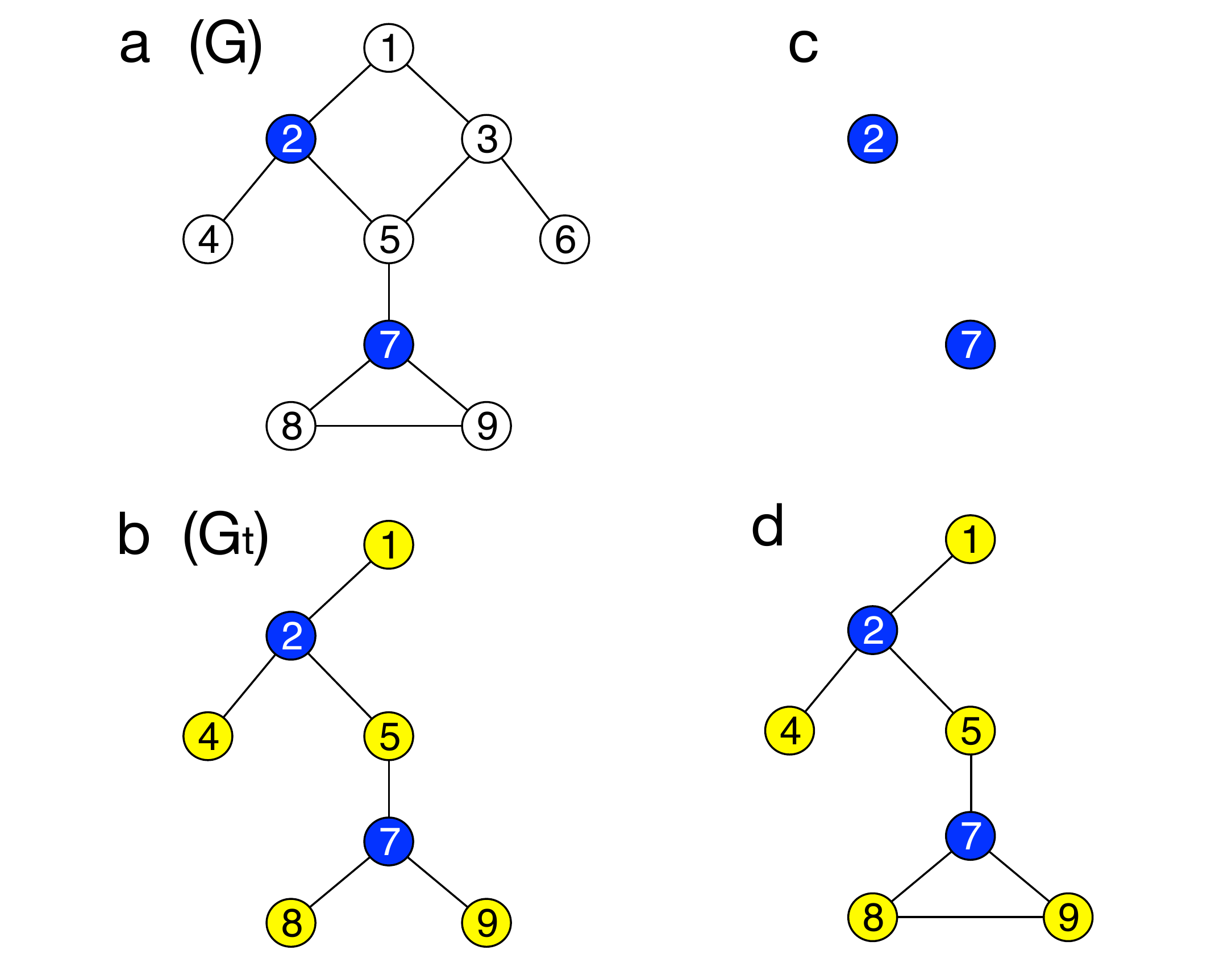}}\\[3pt]  
\label{fig:Gp}
\parbox[c]{15.0cm}{\footnotesize{\bf Fig.~2.} (color online) An illustration of the full network $G$ and the subnetwork $G_t$. (a) The full network $G$ consists of 9 nodes and 10 edges. The subset of nodes to be observed, $V_t$, is $\{2,7\}$ (blue). (b) Subnetwork $G_t$ of network $G$, of which the nodes are composed of $V_t$ (blue) and its adjacent nodes $\partial V_t=\{1,4,5,8,9\}$ (yellow). (c) Subgraph of $V_t$, which consists exclusively of two nodes. (d) Subgraph of $V_t\bigcup \partial V_t$, which adds the edge between nodes 8 and 9 in comparison with the subnetwork $G_t$.}
\end{center}

\section{Methods}
\subsection{Message-passing-based approach}
\subsubsection{Statistical physics analysis}
MDS problem can be mapped to a spin glass model in statistical physics \cite{Zhao2014Statistical}. We follow the same way and give the spin glass model corresponding to t-MDS problem. 

For a given network $G$ and a set $V_t$ of target nodes , we extract 
the subnetwork $G_t$ as defined above, which has $N_t$ nodes. We define a variable $v_i\in\{0,1\}$ on each node of subnetwork $G_t$ to represent the state of the node: it takes the value 1 if node $i$ is occupied and $0$ if not. For a given state configuration $\bm v=(v_1,v_2,\ldots,v_{N_t})$ of $N_t$ nodes, we define an indicator function as follows: 
\begin{equation}
\mathbb I(\bm v)=\prod_{i\in V_t}\big[1-(1-v_i)\prod_{j \in \partial_t i}(1-v_j)\big], 
\end{equation}
where the first product is over all the target nodes, and $\partial_t i$ represents the adjacent nodes of node $i$ in subnetwork $G_t$. $\mathbb I(\bm v)$ takes 1 if $\bm v$ is a t-DS and 0 otherwise. 

We define for each t-DS $\bm v$ an energy function: 
\begin{equation}
E(\bm v)=\sum_{i\in V_t \bigcup \partial V_t}v_i. 
\end{equation} 
The energy gives the number of nodes in t-DS. The corresponding Boltzmann distribution on configurations is 
\begin{equation}
P_{\beta}(\bm v)=\frac{1}{Z(\beta)}e^{-\beta E(\bm v)}\mathbb I(\bm v). 
\end{equation}
Here $\beta>0$ is the inverse temperature and $Z(\beta)$ is the partition function: 
\begin{equation}
\label{eq:partition}
Z(\beta)=\sum_{\bm v}\exp(-\beta \sum_{i\in V_t \bigcup \partial V_t} v_i) \mathbb I(\bm v), 
\end{equation}
where the sum is over all $2^{N_t}$ configurations of $\bm v$. In all configurations, only the t-DS configurations make nonzero contributions to the Boltzman distribution and therefore the distribution only takes the t-DS configurations into consideration. In the large-$\beta$ limit, the distribution is dominated by the ground states, i.e., the t-MDS configurations. 


\subsubsection{Belief-propagation equations}

Model 3 can be solved by using replica-symmetric (RS) mean-field theory \cite{Mezard09,Mezard01}, which gives asymptotically exact results on a large locally tree-like network. A network $G$ has the locally tree-like property if the average length of the shortest cycle going through any node, say $i$, scales as $\log N$ at least, where $N$ is the number of nodes in network $G$. Hence if the original network $G$ is a locally tree-like network, its subnetworks should also be. 

For each pair of connected nodes $(i,j)$ on the subnetwork $G_t$, we introduce two messages $m_{j \to i }^{(v_j,v_i)}$ and $m_{i \to j}^{(v_i,v_j)}$, which are defined on an undirected edge in two opposite directions. 
The expression $m_{j \to i}^{(v_j,v_i)}$ represents the probability that node $j$ takes value $v_j$ and node $i$ takes value $v_i$ in the cavity network where all the edges connecting to node $i$ apart $(i,j)$ are deleted; $m_{i \to j}^{(v_i,v_j)}$ can be obtained by exchanging $i$ and $j$ in $m_{j \to i}^{(v_j,v_i)}$. Consider all the neighbors of node $i$, Bethe-Peierls approximation \cite{Bethe35} assumes that these nodes are independent in the cavity network, and this approximation works well if the underlying network is locally tree-like. 

Recall that the nodes in $G_t$ are divided into two categories: the set of target nodes, $V_t$, and its adjacent nodes $\partial V_t$. Consider a node $j$ in $V_t$ and one of its adjacent nodes $i$. If node $j$ or (and) node $i$ is (are) occupied [$v_j=1$ or (and) $v_i=1$], the node $j$ is observed, and 
thus there are no more constraints on the states of all the neighbors of node $j$ other than $i$. On the contrary if both nodes $j$ and $i$ are empty ($v_i=0$ and $v_j=0$), at least one of adjacent nodes to $j$ apart node $i$ must be occupied to guarantee the observability of node $j$. 
These considerations, together with the Bethe-Peierls approximation, result in the following formulae, called as belief-propagation (BP) equations \cite{Pearl88}, for $m_{j \to i }^{(v_j,v_i)}$\color{black}: 
\begin{equation}
\label{eq:BP1}
m_{j \to i}^{(v_j,v_i)} =\begin{cases}
 \frac{1}{Z_{j \to i}}e^{-\beta}\prod_{k\in \partial_t j \ba i}(m_{k \to j}^{(1,1)}+m_{k \to j}^{(0,1)}), & \text{if}\ (v_j,v_i)=(1,1)\ \text{or} \ (1,0),\\
 \frac{1}{Z_{j \to i}}\prod_{k\in \partial_t j \ba i}(m_{k \to j}^{(1,0)}+m_{k \to j}^{(0,0)}), & \text{if}\ (v_j,v_i)=(0,1),\\
\frac{1}{Z_{j \to i}}\left[\prod_{k\in \partial_t j \ba i}(m_{k \to j}^{(1,0)}+m_{k \to j}^{(0,0)})-\prod_{k\in \partial_t j \ba i}m_{k \to j}^{(0,0)}\right],  & \text{if}\ (v_j,v_i)=(0,0).
\end{cases}
\end{equation}
where $\beta$ is the reverse temperature and $Z_{j \to i}$ is the normalization function: 
\begin{equation}
Z_{j \to i}=2\left[e^{-\beta}\prod_{k\in \partial_t j \ba i}(m_{k \to j}^{(1,1)}+m_{k \to j}^{(0,1)})+\prod_{k\in \partial_t j \ba i}(m_{k \to j}^{(1,0)}+m_{k \to j}^{(0,0)})\right]-\prod_{k\in \partial_t j \ba i}m_{k \to j}^{(0,0)}.
\end{equation}

Note that Eq. (\ref{eq:BP1}) is identical to Eq. (25) in \cite{Zhao2014Statistical}. The difference comes from $\partial V_t$. For a node $j$ in $\partial V_t$, because it does not need to be observed, the term $\prod_{k\in \partial_t j \ba i}m_{k \to j}^{(0,0)}$ in the expression of $m_{j \to i}^{(0,0)}$ can be deleted, and thus the new $m_{j \to i }^{(0,0)}$ is equal to $m_{j \to i }^{(0,1)}$. Hence the BP equations of messages $m_{j \to i}^{(v_j,v_i)}$ can be simplified as  
\begin{equation}
\label{eq:BP2}
m_{j \to i}^{(v_j,v_i)}=\frac{e^{-\beta v_j}\prod_{k\in\partial_t j\ba i}(m_{k\to j}^{(1,v_j)}+m_{k\to j}^{(0,v_j)})}{2\left[e^{-\beta}\prod_{k\in \partial_t j \ba i}(m_{k \to j}^{(1,1)}+m_{k \to j}^{(0,1)})+\prod_{k\in \partial_t j \ba i}(m_{k \to j}^{(1,0)}+m_{k \to j}^{(0,0)})\right]}, \ \ \ \text{for}\ j\in \partial V_t.
\end{equation}

Once a solution of BP equations (\ref{eq:BP1}) and (\ref{eq:BP2}) is found, the marginal probability $m_{i}^{(1)}$ of node $i$'s state being 1 is estimated as 
\begin{equation}
\label{eq:mar}
m_{i}^{(1)} =
\frac{e^{-\beta}\prod_{j \in \partial_t i}(m_{j \to i}^{(0,1)}+m_{j \to i}^{(1,1)})}{e^{-\beta}\prod_{j \in \partial_t i}(m_{j \to i}^{(0,1)}+m_{j \to i}^{(1,1)})+\prod_{j \in \partial_t i}(m_{j \to i}^{(1,0)}+m_{j \to i}^{(0,0)})-\chi_{V_t}(i)\prod_{j \in \partial_t i}m_{j \to i}^{(0,0)}},
\end{equation}
where $\chi_A(i)$ is an indicator function, which takes 1 if $i\in A$ and 0 otherwise. 


\subsubsection{BPD algorithm}
Now we construct the belief-propagation-decimation (BPD) algorithm for the t-MDS problem. To distinguish it from the normal BPD algorithm for the MDS problem, proposed in \cite{Zhao2014Statistical}, we call the solving algorithm the target belief-propagation-decimation (t-BPD) algorithm. 

For a given network $G$ and a subset $V_t$ of target nodes to observe, the t-BPD algorithm first extracts the subnetwork $G_t$ induced by $V_t$ and $\partial V_t$. The t-DS $\Gamma$ starts from an empty set. We initialize the messages $m_{j\to i}^{(v_j,v_i)}$ and $m_{i\to j}^{(v_i,v_j)}$ for each edge $(i,j)$ in $G_t$ to be uniform distribution: $m_{j\to i}^{(v_j,v_i)}=m_{i\to j}^{(v_i,v_j)}=0.25$. The inverse temperature $\beta$ is set to be a large value, e.g., 10.  The messages are updated according to BP equations [Eqs. (5) and (7)] for a number, say 50, of rounds. During each round, we examine each node in the current network. When node $j$ is examined, the outgoing messages $m_{j \to i}^{(v_j,v_i)}$ to all its neighbors $i \in \partial_t j$ are updated according to Eq. (5) if $j\in V_t$ and to Eq. (7) if $j \in \partial V_t$.  After each round of the update process, a small fraction (e.g., 1$\%$) of unoccupied nodes in $G_t$ that have the largest occupation probability $m_i^{(1)}$ are added to the set $\Gamma$ and, because their states have been determined, the messages coming from and into them need not to be recomputed. 

These newly occupied nodes cause their unobserved neighbors in $V_t$ to be observed. Once an unoccupied node $j$ in $V_t$ is observed, it does not impose any constraint on the state of its neighbor, say $i$, and, in this scenario, $m_{j \to i}^{(0,1)}=m_{j \to i}^{(0,0)}$ and thus the messages sent from node $j$ are updated according to Eq. (7). The above process is repeated until all the nodes in $V_t$ are observed and the final $\Gamma$ is the estimated t-MDS. 

The main differences between t-BPD algorithm and BPD algorithm in \cite{Zhao2014Statistical} are (1) t-BPD algorithm adds an extra step to extract the subnetwork $G_t$, and (2) all the messages obey the identical BP equations in BPD algorithm, whereas the messages obey the two distinct sets of BP equations in t-BPD algorithm, depending the type of the node ($\in V_t$ or $\in \partial V_t$) where the messages come from.  The pseudo code of the t-BPD algorithm is presented in Appendix A. 

\subsection{Local algorithms}
We consider two local algorithms, the greedy algorithm and the local-consensus algorithm, which have been shown to have comparative satisfactory performances in solving the MDS problem on both random-generated networks and real-world networks \cite{Zhao2014Statistical,Sun2016}. We generalize these two algorithms to the t-MDS problem. 

Both of the two algorithms use the concept of impact of a node. The impact of an unoccupied node $i$, denoted as $I(i)$, refers to the number of nodes that will be newly observed if $i$ is occupied. 
For the t-MDS problem, the unoccupied nodes in subnetwork $G_t$ can be divided into three types: (1) the observed nodes in $V_t$, (2) the unobserved nodes in $V_t$, and (3) the nodes in $\partial  V_t$. For a node $i$ of the type (1) or (3), its impact $I(i)$ is equal to the number of its unobserved neighbors in $V_t$; and for a node $i$ of type (2), its impact $I(i)$  is equal to the number of its unobserved neighbors in $V_t$ plus 1.          
\subsubsection{Greedy algorithm}
For the t-MDS problem, the greedy algorithm is based on the same idea as that in the MDS problem, that is, repeatedly occupying the node that has the largest impact in the current network. Specifically, it first extracts the subnetwork $G_t$ induced by set of nodes, $V_t$, and its neighbors $\partial V_t$. Initially, the t-DS  is empty. For a node $i\in \partial V_t$, $I(i)$ is its degree in $G_t$, i.e., the number of its adjacent nodes in $V_t$; and for a node $i \in V_t$, $I(i)$ is its degree in $G_t$ (also in $G$) plus 1. In each step, the greedy algorithm adds a node $i$ randomly from the set of unoccupied nodes with the maximal impact to the current t-DS. All the adjacent nodes of $i$ are observed. Node $i$ is removed along with all the edges connecting to $i$ from the current network. The remaining nodes, if there are any, update their values of impact. This process is repeated until all the nodes in $V_t$ are observed. We call this greedy algorithm the target greedy impact (t-GI) algorithm. If the subnetwork $G_t$ has maximum "modified" degree $\Delta_t$, it can be proved that the size of t-DS found by t-GI algorithm is at most $log(\Delta_t)$ times of the size of t-MDS. More details are provided in Appendix B. 

\subsubsection{Local-consensus algorithm}
The implementation of the t-BPD and t-GI algorithms both need a central planner who has complete information of subnetwork $G_t$ and each node's impact. In reality, however, some social and economical networks are massive in sizes and complexity, such that it is difficult to obtain the complete network information, and hence utilizing the t-BPD and t-GI algorithms may be infeasible for this scenario. To solve this problem, we need a decentralized algorithm. For the MDS problem, the first author and a collaborator proposed a decentralized local-consensus algorithm, which can lead to a DS with size approaching the smallest possible value for both random networks and real-world networks \cite{Sun2016}. We now generalize this algorithm to the t-MDS problem. To highlight the target observation, we call the solving algorithm the target local-consensus (t-LC) algorithm. 

Similarly to the t-GI algorithm, the t-LC algorithm first extracts a subnetwork $G_t$. We assume that each node in $G_t$ only knows the latest impact values and occupation states (being occupied or not) of its neighbors. In each step, each unoccupied node $i$ checks the states of its neighbors and makes decision. Specifically, if node $i\in V_t$ and is unobserved, it will suggest occupying an adjacent node $j$ if and only if $j$ has the largest impact among $i$'s neighbors and also $I(j)$ is not less than $I(i)$. Instead, if node $i \in V_t$ and is observed or node $i \in \partial V_t$, it will suggest occupying an adjacent unobserved node $j$ in $V_t$ as long as $I(j)$ is not less than $I(i)$. Then, we determine the candidate set $\Omega$ of nodes to occupy. An unobserved node $j$ in $V_t$ enters into the candidate set $\Omega$ only if all of its neighbors recommend occupying it. An unoccupied node $j$ in $\partial V_t$ or an observed but unoccupied node $j$ in $V_t$ enters into the candidate set if only all of its unobserved neighbors in $V_t$ recommend to occupy it. A node, say $k$, is selected randomly from the current candidate set $\Omega$ and is occupied.  All of $k$'s neighbors in $V_t$ are thus observed. Each unoccupied node in the current network updates its impact value. The process is repeated until all the nodes in $V_t$ are observed. 

\section{Results}
Now we discuss the performances of the above three algorithms (t-BPD, t-GI, and t-LC) for the t-MDS problem on simulated networks and real-world networks. Here we consider two mechanisms for selecting the target nodes that need to be observed: 1. random selection, where a fraction $f$ of the nodes are selected randomly from the network, and 2. snowball selection, where a node, say $i$, is selected randomly at first, then all of its neighbors are also recruited, and the process is repeated until the number of selected nodes reaches $fN$, where $N$ is the number of nodes in the network. In the snowball selection mechanism, the selected nodes form a connected component. 

We first evaluate the algorithms on three types of random networks: Erd\"{o}s-R\'{e}nyi (ER), regular random (RR), and scale-free (SF) networks. 
In ER network, the degree of nodes is distributed according to a Poisson distribution; and in RR network, each node has the same degree.  
The SF networks used in this work are generated via the static model \cite{Goh}, which is a common used generation model of SF networks. 
In the static model, node $i$ is assigned to a weight $p_i=\frac{i^{-\xi}}{\sum_{j=1}^Nj^{\xi}}$. Two nodes $i,j$ are selected independently with probabilities as $p_i$ and $p_j$. If there does not exist an edge between $i$ and $j$, then an edge $(i,j)$ is added. The final network has a power-law degree distribution $P(k)\propto k^{-\gamma}$ for $k \gg1$, with decay exponent $\gamma=1+\frac{1}{\xi}$ ($0\leq\xi<1$). Throughout this paper, $\xi$ is set to be $0.5$ and thus $\gamma=3$. 

We show the proportion $n_0$ of occupied nodes (first row) and proportion $n_1$ of observed nodes (second row) for varying fraction $f$ of target nodes on random networks with a fixed mean degree  $\langle K\rangle=10$ in Fig.~3 and that for networks with different mean degree  $\langle K \rangle$ while fixing the fraction $f$ of target nodes to be $0.5$ in Fig.~4. Note that the observed nodes here refer to all the nodes in full network that are observed by the t-DS. In general, we find that the snowball mechanism needs to occupy fewer nodes compared to the random mechanism, and, in the meantime, fewer nodes can be observed than that in the random mechanism. The reason is that, compared to snowball mechanism, the nodes selected through random mechanism are typically scattered across the network, which leads to a lower average number of connections among target nodes and more adjacent nodes outside the set of target nodes. The t-DS obtained by t-BPD algorithm is always the smallest one. The size of t-DS identified by t-GI and t-LC are close to each other. Interestingly, we find that the t-DS detected by t-GI are able to observe more nodes than that of the other two algorithms, in particular, under the random selection mechanism. This indicates that the t-GI can construct a t-DS with stronger observation capacity.

\begin{center}
\resizebox{0.8\textwidth}{!}
{\includegraphics{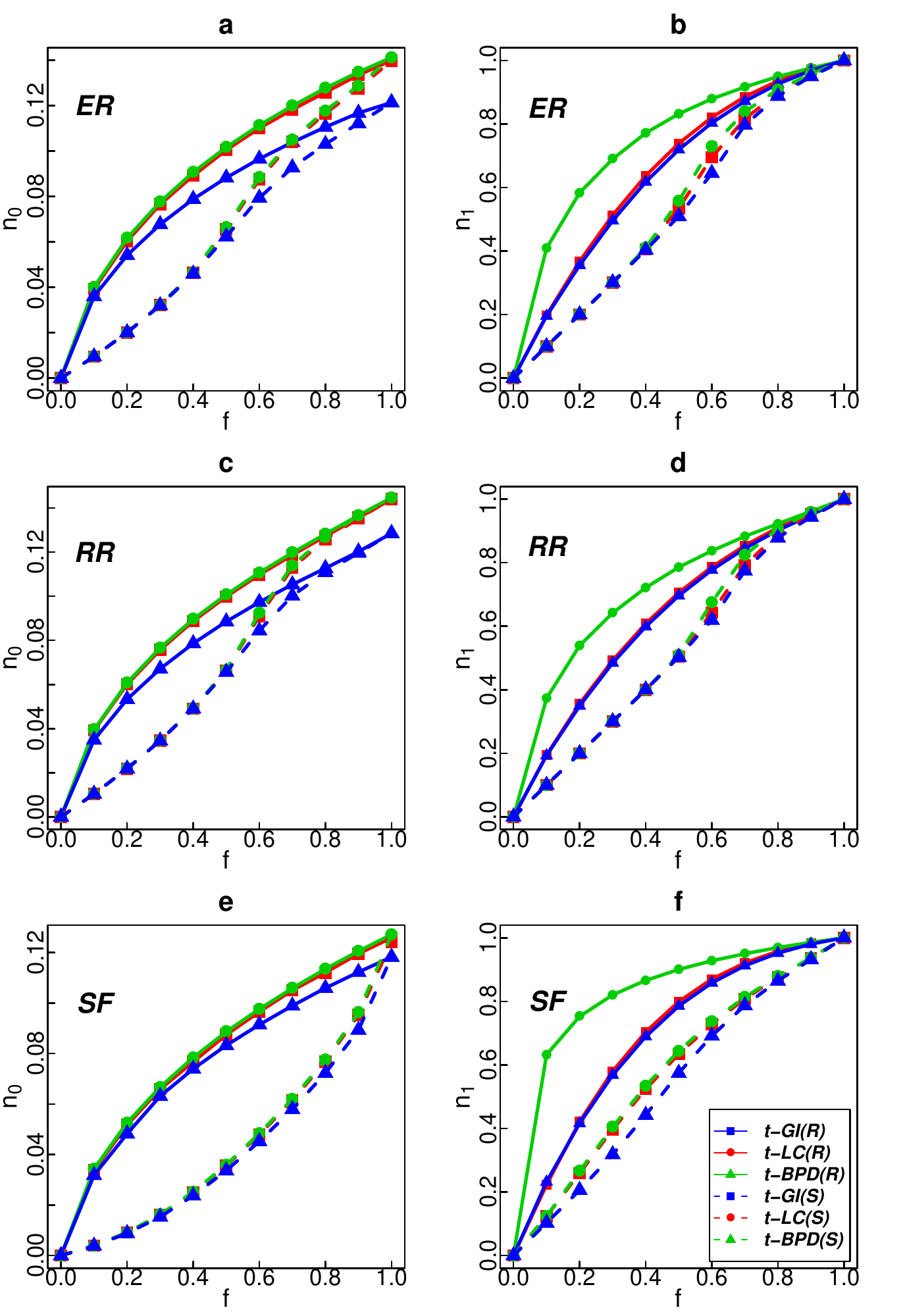}}\\[3pt]  
\parbox[c]{15.0cm}{\footnotesize{\bf Fig.~3.} (color online) Target observation of random networks with mean degree  
$\langle K \rangle=10$. In the first row, we show the fraction $n_0$ of nodes in the t-DS obtained by t-GI(green), t-LC(red) and t-BPD(blue) as a function of the fraction $f$ of target nodes; in the second column, we show the fraction $n_1$ of nodes observed by the t-DS obtained by t-GI, t-LC, and t-BPD. We consider three types of random networks: (a) and (b) ER, (c) and (d) RR, and (e) and (f) SF. In each subfigure, the solid line describes the random selection mechanism and the dashed line describes the snowball selection mechanism. Each data point is the averaged result over 48 network instances with $N = 10^5$ nodes. The standard deviations of results among network instances are smaller than the symbol size and are thus not shown here.}
\end{center}

\begin{center}
\resizebox{0.8\textwidth}{!}
{\includegraphics{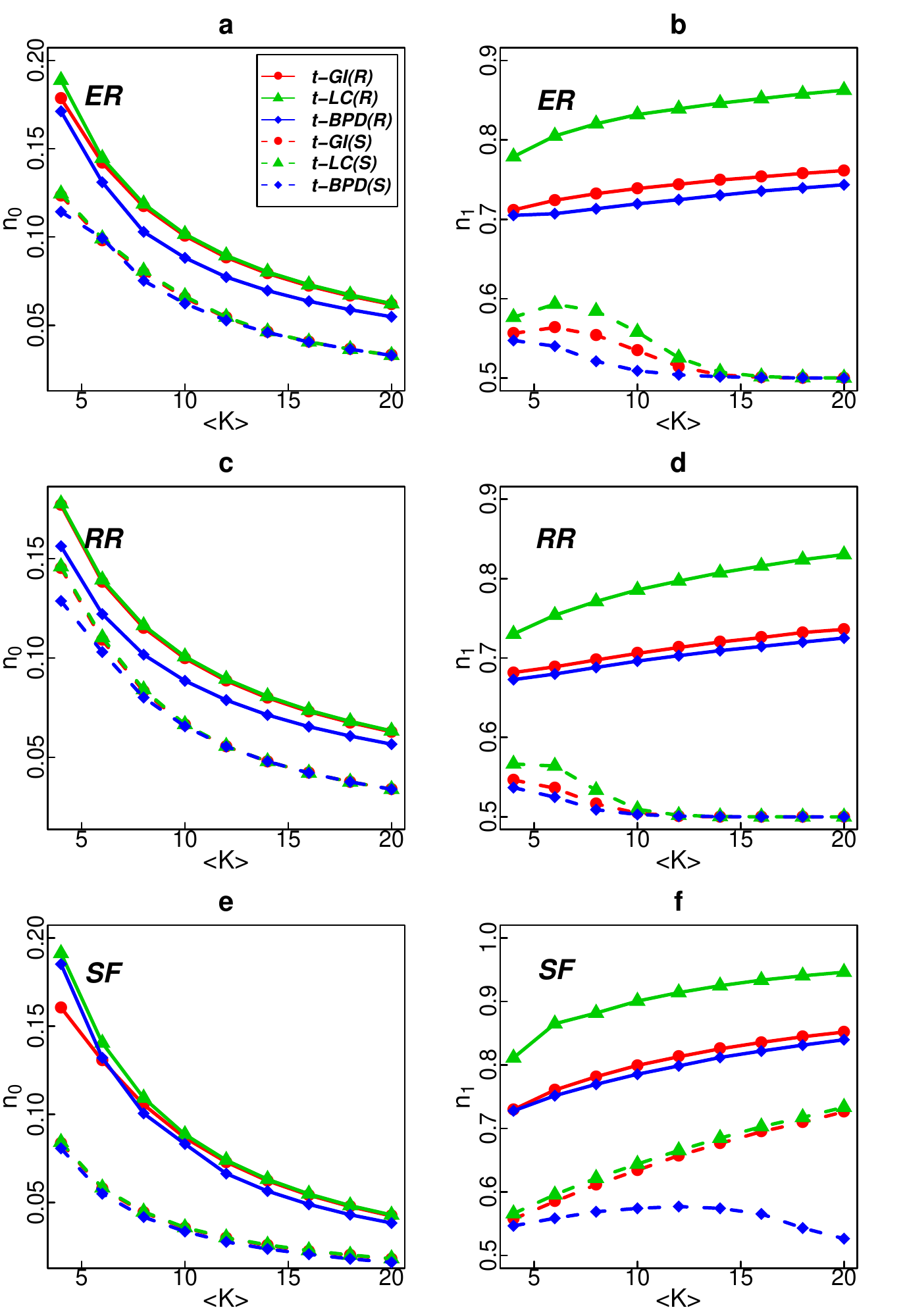}}\\[3pt]  
\parbox[c]{15.0cm}{\footnotesize{\bf Fig.~4.} (color online) Target observation of random networks with a fixed fraction $f=0.5$ of target nodes. The subfigures in the first row show the fraction $n_0$ of nodes in the t-DS obtained by t-GI(green), t-LC(red) and t-BPD(blue) as a function of the mean degree  $\langle K \rangle$ of the networks; the subfigures in the second column show the fraction $n_1$ of nodes observed by the t-DS obtained by t-GI, t-LC and t-BPD. In each subfigure, the solid line describes the random selection mechanism and the dashed line describes the snowball selection mechanism. All results are averaged over 48 independent network instances with $N = 10^5$ nodes. The standard deviations of results among network instances are smaller than the symbol size and are thus not shown here.}
\end{center}

We also compare the time complexity of the three algorithms. Both t-LC and t-GI algorithms occupy a node in each step and then update the impact value for each remaining node. If the size of set $V_t$ of target nodes is of the same order as the size $N$ of network , the running times of t-LC and t-GI are proportional to $N^2$. When the t-BPD algorithm is applied as a solver, it in each step occupies a tiny fraction ($1\%$) of the nodes in current network and then updates the message along each remaining edge. If the network is sparse, i.e., the number $N$ of nodes is of the same order as the number $M$ of edges, the running time of t-BPD is proportional to $N\log N$. Fig.~5 presents the running times (in seconds) of the three algorithms on SF network instances of size $N$ ranging from $10^4$ to $10^6$. In general, we find that t-LC and t-GI are faster than t-BPD on the small- and medium-scale networks ($N<10^5$) , but the advantages of the two local algorithms in running time become smaller with the increase of the size of network. The running times of two local algorithms are of the same order, and the t-LC algorithm performs slightly faster than the t-GI algorithm. A possible reason is that the t-GI algorithm involves an additional sort operation in comparison with the t-LC algorithm. We also compare two selection mechanisms: random and snowball, of the target nodes. The results indicate that the selection mechanisms do affect the efficiencies of the proposed algorithms. The two local algorithms perform faster under the snowball mechanism. This can be explained as follows. The snowball mechanism requires fewer occupied nodes when compared with the random mechanism. Hence the two local algorithms have less number of iterations under the snowball mechanism. In contrary, the t-BPD algorithm runs faster under the random mechanism. The main reason is that the random mechanism constructs a sparser subnetwork, i.e., the subnetwork has fewer edges, than the snowball mechanism does, which suggests that fewer messages need to be updated in each step.     

\begin{center}
\resizebox{0.7\textwidth}{!}
{\includegraphics{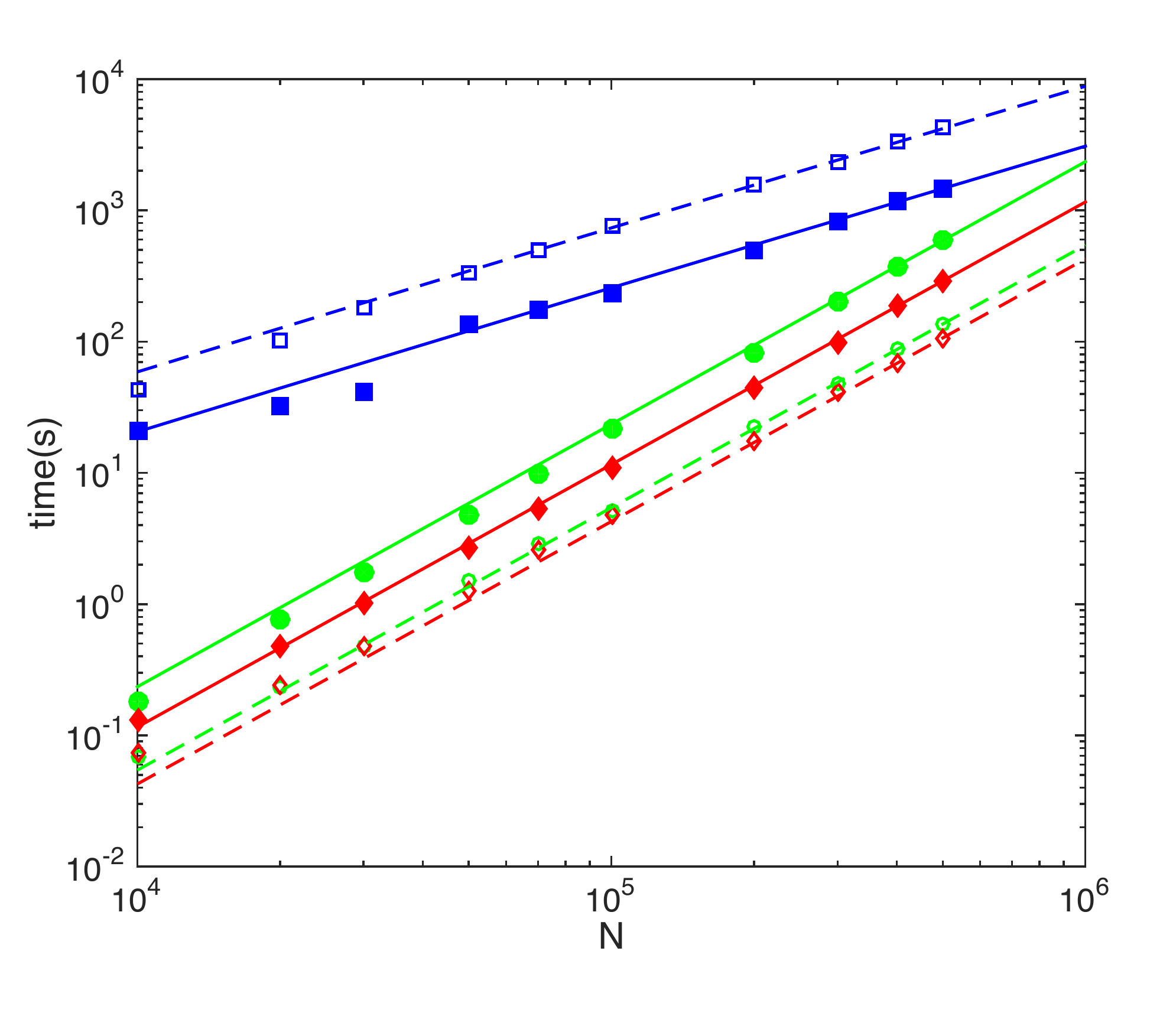}}\\[3pt]  
\parbox[c]{15.0cm}{\footnotesize{\bf Fig.~5.} (color online) Running times (in seconds) of the three algorithms on SF network instances of mean degree  $\langle K \rangle=10$ and size $N$. The results obtained by t-BPD, t-GI, and t-LC are shown as square symbols, circle symbols and diamond symbols, respectively. The full symbols represent the random selection mechanism, and the empty symbols represent the snowball selection mechanism. The solid and dashed lines are the fitting curves of the running time under random and snowball selection mechanism, respectively. The fitting curves for t-BPD running time are of the form $aN\log_{10}N$, with parameter $a=7.25\times 10^{-5}$ (for random mechanism) and $a=1.12\times 10^{-4}$ (for snowball mechanism). The fitting curves for two local algorithms are both of the form $bN^2$. Specifically, for the t-GI algorithm, the parameter $b=2.35\times 10^{-9}$ (for random mechanism) and $b=5.44\times 10^{-10}$ (for snowball mechanism); and for the t-LC algorithm, the parameter $b=1.16\times 10^{-9}$ (for random mechanism) and $b=4.26\times 10^{-10}$ (for snowball mechanism).}
\end{center} 

As mentioned in Section 1, a natural solution for the t-MDS problem is to construct a near-minimum DS by using some algorithm first, and then remove the nodes from the DS whose removal will not make any target nodes unobservable. We call this a type I alternative approach. To compare the type I alternative approach with the proposed approach, we adopt BPD, GI and LC algorithms, respectively, to get a near-optimal DS and then remove the redundant nodes. We compute the relative difference of the size of the t-DS found by the proposed approach and that obtained by the alternative approach: $\frac{\Delta n_0}{n_0}=\frac{n_0(alternative)-n_0(proposed)}{n_0(proposed)}$. Fig.~6 (a) presents the boxplots\footnote{The boxplot is a method to summarize and display the distribution of a set of continuous data in a compact way. It is constructed to highlight five important statistical characteristics of data: the median, the first and third quartiles, and the minimum and maximum. The outliers, that is the data points that are far outside the rand of the other data points, are plotted as individual points.} of relative difference over random networks with mean degree  $\langle K \rangle \in\{4,6,\ldots,20\}$, where the fraction $f$ of target nodes is fixed to be 0.5. We observe that the proposed approach can indeed find a smaller t-DS than the type I alternative approach does. In general, the snowball mechanism has a larger  relative difference $\frac{\Delta n_0}{n_0}$ than the random mechanism. The results suggest that the optimum DS to full observation is not the optimum one in terms of target observation. In addition, we also compare the proposed approach with a type II alternative approach, that is, applying BPD, GI, and LC algorithms for the subnetwork composed of only target nodes and edges between them. As expected, the proposed approach has a superior performance in solving the t-MDS problem [Fig.~6 (b)].  

\begin{center}
\resizebox{1\textwidth}{!}
{\includegraphics{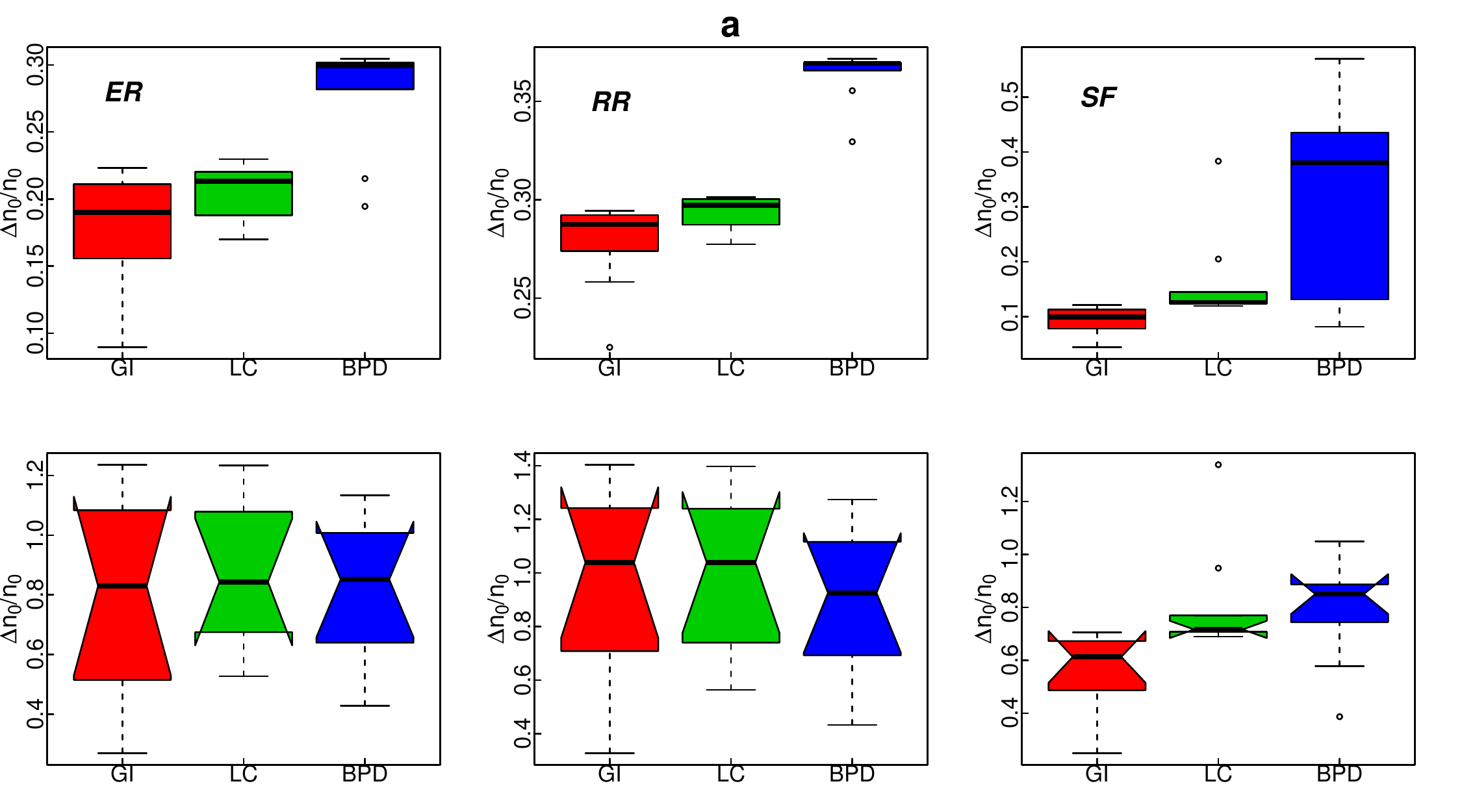}}\\[3pt]
\resizebox{1\textwidth}{!}
{\includegraphics{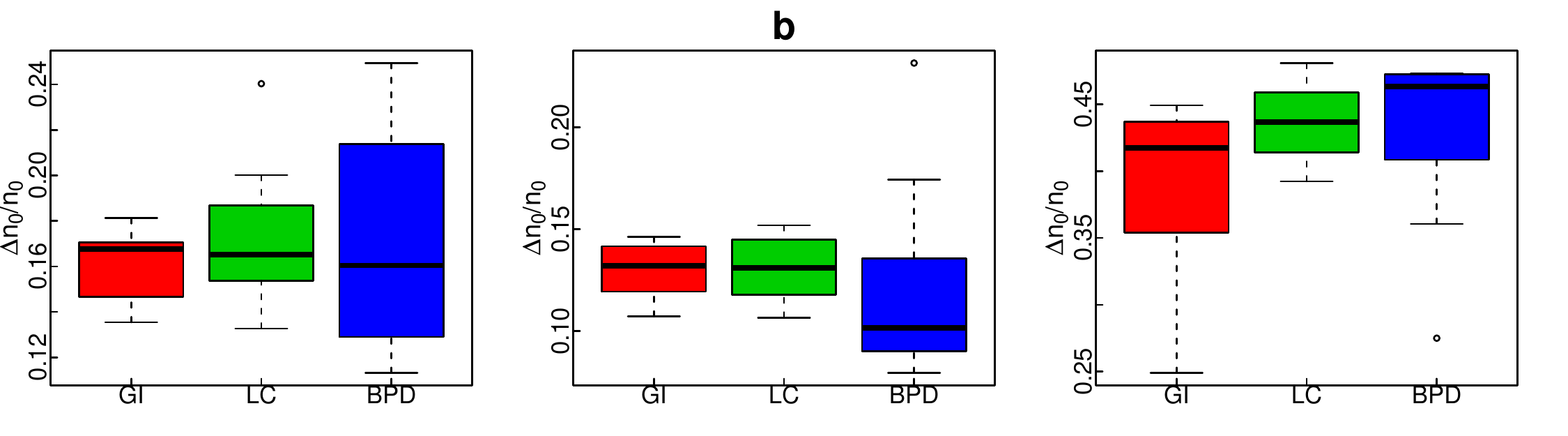}}\\[3pt]   
\parbox[c]{15.0cm}{\footnotesize{\bf Fig.~6.} (color online) Comparison of performances of the proposed approaches and alternative approaches. We consider three types of random networks: ER (left panel), RR (middle panel), and SF (right panel), with mean degree  $\langle K \rangle\in \{4,6,\ldots,20\}$. Each network instance has $N=10^5$ nodes. (a) The boxplots of the relative difference $\frac{\Delta n_0}{n_0}$ of the size of t-DS found by the proposed approaches and that by the type I alternative approach over networks with distinct mean degree for the random selection mechanism (first row) and the snowball selection mechanism (second row). (b) The boxplots of the relative difference $\frac{\Delta n_0}{n_0}$ of the size of t-DS found by the proposed approaches and that by the type II alternative approach over networks with distinct mean degree for the random selection mechanism. In each subfigure, the outliers are plotted as individual points.}
\end{center}

In the end, we apply the three algorithms to several real-world networks. The structural characteristics of the networks
are given in Appendix C. Fig.~7 summarizes the simulation results. We again find that the snowball selection mechanism requires fewer occupied nodes. Among the three algorithms, although t-BPD still gives the t-DS with minimum size, the two local algorithms, t-GI and t-LC, perform almost as well as the t-BPD one for some of these examined network instances. For the random selection mechanism, the t-DS found by the t-GI algorithm can observe noticeably more nodes than the other two algorithms.  

To further explore the performances of three algorithms on the real-world networks, it is worthy and interesting to investigate the structural characteristics of the t-DS identified by each algorithm. We find that, for each real-world network used in this paper, the t-DSs detected by the three algorithms have similar structures. Take the Loc network \cite{Cho} as an example (see more information of this network in Appendix C), the degree distribution of the nodes in t-DS is same as the degree distribution of all nodes in the original network, both of which are power-law distributions (Fig.~8). Note that the degree of a node $i$ in t-DS refers to the degree of node $i$ in the original network. This fact suggests that there is indeed a significant number of low-degree nodes in the t-DS found by each algorithm. The large number of connected components of t-DS indicates that the nodes in t-DS tend to form a lot of scattered clusters (Tab. 1). In term of the selection mechanisms of the target nodes, the snowball mechanism always yields a denser t-DS which forms few components and has a larger mean degree when compared with the random mechanism. 

\section{Conclusion}
In this work, we study the target observation of networks and propose the t-MDS problem, i.e., occupying as few as possible nodes such that a specified set of nodes is observed, which can be seen as a generalization of classical MDS problem. We develop t-BPD (a message-passing algorithm), t-GI, and t-CI (two local algorithms) to identify an approximately minimum t-DS. We investigate two selection mechanisms--random and snowball--to select the set of target nodes. Extensive numerical experiments demonstrate that the three proposed algorithms are efficient in finding the t-DS for target observation on random networks and real-world networks. In general, the snowball mechanism needs fewer occupied nodes than the random one. The t-BPD algorithm can find a smaller t-DS compared with the two local algorithms, while the t-DS detected by the t-GI algorithm can observe the largest number of nodes. Further progress can be achieved by extending the proposed methods to multiplex networks.

\begin{center}
\resizebox{0.8\textwidth}{!}
{\includegraphics{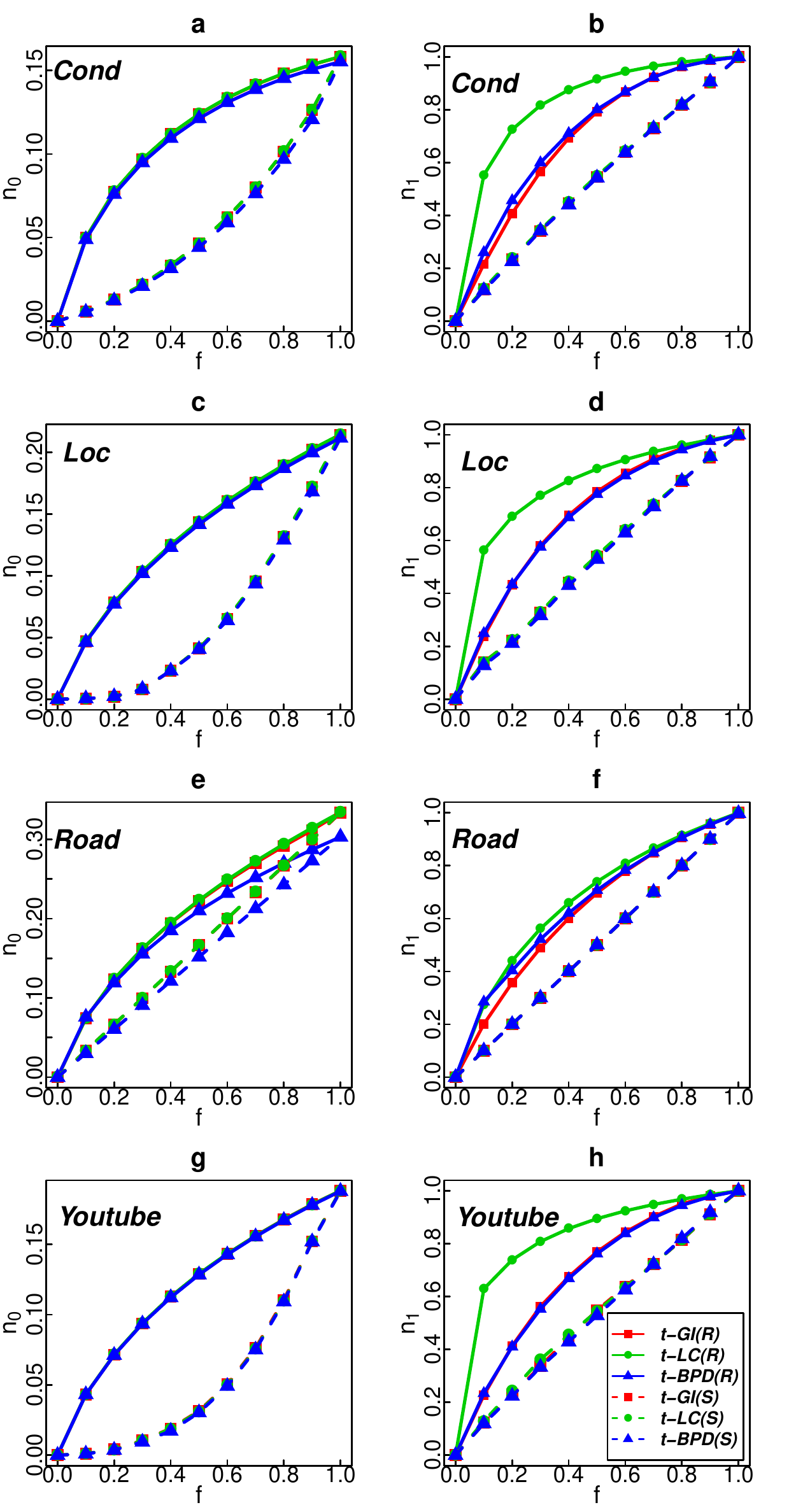}}\\[3pt]  
\parbox[c]{15.0cm}{\footnotesize{\bf Fig.~7.} (color online) Target observation of four real-world networks. For each network, we show the fraction $n_0$ of occupied nodes (first row) and the fraction $n_1$ of nodes observed by the occupied nodes (second row) as a function of the fraction $f$ of target nodes. In each subfigure, the solid line describes the random selection mechanism and the dashed line describes the snowball selection mechanism.}
\end{center}


\begin{center}
\resizebox{1\textwidth}{!}
{\includegraphics{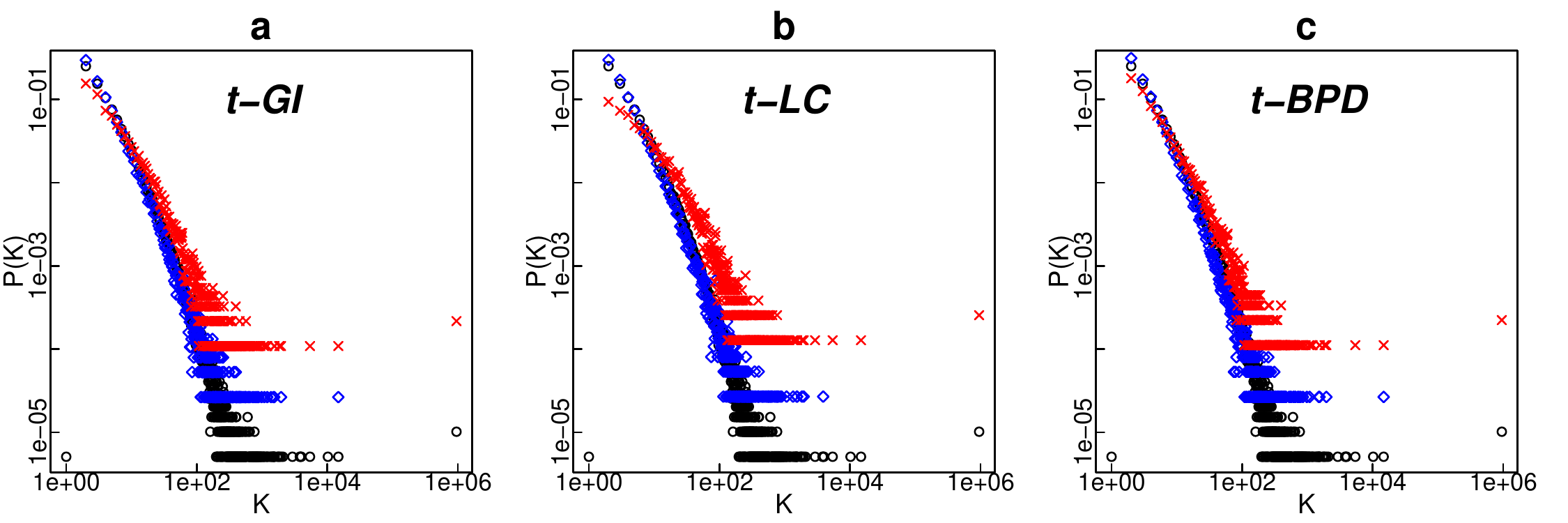}}\\[3pt]  
\parbox[c]{15.0cm}{\footnotesize{\bf Fig.~8.} (color online) Degree distribution of the nodes in t-DS and that of all nodes in Loc network (in log-log plot). The fraction $f$ of target nodes is set to be 0.5. The black circles represent the degree distribution of all nodes, and the blue diamonds the occupied nodes in t-DS under the random mechanism, and the red crosses the occupied nodes in t-DS under the snowball mechanism. }
\end{center}

\begin{table}[htbp]
\centering
\caption{Structural statistics of t-DS identified by t-GI, t-LC, and t-BPD algorithms on Loc network. The fraction $f$ of target nodes is set to be 0.5. For each t-DS, we list the number of nodes, $N$, the number of connected components, $N_{cc}$, the mean degree , $\langle K\rangle$, the decay exponent of degree distribution, $\gamma$. For comparison, we also list the statistics of the original network.}
   \begin{tabular}{lccccccc}
   \toprule
   &\multicolumn{6}{c}{\ \ \ \ \ \ \ \ \ \ \ \ \ \ \ \ \  \ \ \ \ Target dominating set}\\
   \cmidrule(r){3-8} 
   & Network & \multicolumn{3}{c}{Random} & \multicolumn{3}{c}{Snowball}\\
   \cmidrule(r){3-5} \cmidrule(l){6-8}
   && GI & LC & BPD & GI & LC & BPD\\
   \midrule 
   $N$ &196591 & 37781 & 37728 & 37545 & 8191 & 8079 & 7942\\
   $N_{cc}$ & 1 & 26344 & 25851 & 30665 & 1426 & 1106 & 2812\\ 
   $\langle K\rangle$ & 9.668 & 4.269 & 4.462 & 5.358 & 114.562 & 130.235 & 130.471 \\
   $\gamma$ & 2.784 & 2.631 & 2.672 & 2.664 & 1.785 & 1.805 & 1.717 \\
   \bottomrule
   \end{tabular}
\end{table}

\section*{Acknowledgements}
This work is supported by the National Natural Science Foundation of China (No.11605288) and Fund for building world-class universities (disciplines) of Renmin University of China. 

\section *{Appendix A. T-BPD algorithm}
\begin{algorithm}
\caption{Target Belief-Propagation-Guided Decimation algorithm}
\label{alg:1}
\begin{algorithmic}
\State \bf{Input}: \rm network $G$, a subset $V_t$ of target nodes, maximum iteration number $T_0$, and inverse temperature $\beta=10$.
\State \bf{Initialization:} \rm Extract the subnetwork $G_t$, composed of $V_t$ and $\partial V_t$ as nodes and linkages connecting with at least one node in $V_t$ as edges. Set $m_{j \to i}^{(v_j,v_i)}=0.25$ for any edges $(i,j)$ in $G_t$ and the t-DS $\Gamma=\emptyset$
\Repeat
\State /*update messages*/
\For{$T=1$ to $T_0$}
\For  {each unobserved node $j$ in $V_t$}
  \State update $m_{j \to i}$ according to equation (\ref{eq:BP1}), for each adjacent node $i$;
\EndFor
\For {each node $j$ in $\partial V_t$ and each observed but unoccupied node $j$ in $V_t$}
  \State update $m_{j \to i}$ according to equation (\ref{eq:BP2}), for each adjacent node $i$;
\EndFor
\EndFor
\State /* occupy nodes */
\For {each unoccupied node $i$}
\State   compute $m_{i}^{(1)}$ according to equation (\ref{eq:mar})
\EndFor
\State choose a fraction of nodes (e.g., 1$\%$ of unoccupied nodes in $G_t$) that have the largest occupation probability $m_i^{(1)}$; occupy these nodes and then update the states of their unobserved neighbors in $V_t$ to be observed;
\State /* Decimation process */
\State add the newly occupied nodes to set $\Gamma$; remove the newly occupied nodes and all the edges connecting to them from the current network;
\Until{all the nodes in $V_t$ are observed.}
\State \bf{Output}\rm: the t-DS $\Gamma$. 
\end{algorithmic}
\end{algorithm}

\section *{Appendix B. Approximation ratio of the t-GI algorithm}
For the standard MDS problem, it is well known that the greedy algorithm can provide an $O(\log\Delta)$-approximation of the size of MDS, where $\Delta$ is the maximum degree of the original network $G$. The key point is utilizing the equivalence between the MDS problem and the set cover (SC) problem. Since any MDS problem can be converted into a SC problem, the results for the greedy algorithm on the SC problem can be specialized to the MDS problem, and thus yields the $O(\log\Delta)$ approximation ratio. 

We now show that the t-MDS problem can also be formulated as a SC problem. Given an undirected network $G=(V,E)$ and a set of target nodes $V_t\subset V$, we extract a subnetwork $G_t$ as the nodes in $V_t$ and $\partial V_t$ along with the edges that links at least one node in $V_t$.  A SC instance $(U,\mathcal{S})$ is constructed as follows: the universe $U$ is $V_t$ and the family of subsets is $\mathcal{S}=S1\bigcup S2$, where $\mathcal{S}_1=\{S_i|i\in V_t\}$ such that $S_i$ consists of the node $i$ and all its adjacent nodes in $V_t$, and $\mathcal{S}_2=\{S_i|i\in \partial V_t\}$ such that $S_i$ consists of all adjacent nodes to $i$ in $G_t$.  

If $\Gamma_t$ is a t-DS for $G_t$ (also for $G$), then $C_t=\{S_i|i\in\Gamma\}$ is a solution for the SC problem and vice verse. Hence the size of t-MDS for $G$ equals the size of a minimum SC for $(U,\mathcal{S})$. We define a modified degree of node $i\in V_t\bigcup \partial V_t$  as the size of its corresponding set $S_i\in \mathcal{S}$. Denote the maximum modified degree of $G_t$ as $\Delta_t$. According to the results given by Chvatal \cite{Chvatal}, we conclude that the t-GI algorithm can find an $O(\log \Delta_t)$-approximation of a t-MDS.  

\section *{Appendix C. Real-world network instances}
The structural statistics of these networks are listed in Tab.~C.1. 
\begin{center}
{\footnotesize{\bf Table C.1.} Statistics of four real-world network instances used in Fig.~7. For each network, we list the number of nodes, $N$, the number of edges, $M$, the mean degree, $\langle K \rangle$, the diameter, $D$, and the average clustering coefficient, $CC$.
\\
\vspace{2mm}
\begin{tabular}{cccccc}
\hline
{Network}      & {$N$}       & {$M$}    &{$\langle K \rangle$}  &{$D$}   &{$CC$}\\\hline
{Cond \cite{Jure2007Graph}}           & {23 133}     &{93 497} &{43.69}  & {14} &   {0.6334} \\
{Loc \cite{Cho}}              & {196 591}   & {950 327} &{9.668} &{14} & {0.2367} \\
{Road \cite{Jure2009Community}}   & {1 088 092}        & {1 541 898}  &{2.834} &{786} &{0.0465} \\
{Youtube \cite{Yang2012Defining}} & {1 134 890}   & {2 987 624} &{5.265} &{20} &{0.0808} \\
\hline
\end{tabular}}
\end{center}


\end{document}